\documentstyle[epsf,11pt]{article}
\def\baselinestretch{1.1}
\title{\bf One way to characterize the compact structures of 
lattice protein model\thanks{This project was supported partly by Chinese 
Natural Science Foundation.} 
}
\author{Bin Wang$^1$, Zu-Guo Yu$^{2,1}$\\   
 {\small $^1$Institute of Theoretical Physics, Chinese Academy of Sciences},\\
  {\small P.O. Box 2735, Beijing 100080, P. R. China.}\\
  {\small $^2$Department of Mathematics, Xiangtan Universiy,}\\
   {\small Hunan 411105, P.R. China}
  }
\begin{document}
\maketitle
\begin{abstract}

On the study of protein folding, our understanding about the protein 
structures is limited. In this paper we find one way to characterize the 
compact structures of lattice protein model. A quantity called {\bf Partnum} 
is given to each compact structure. The Partnum is compared with the 
concept {\bf Designability} of protein structures emerged recently. It is shown that the highly
designable structures have, on average, an atypical number of local 
degree of freedom. The statistical property of Partnum and its 
dependence on sequence length is also studied.
\end{abstract}

\section{Introduction}
   The study of protein folding is fundamental on both theory and 
application. In order to tackle protein folding problem physically, 
it is important to pay much attention to concrete proteins and consider 
the details of interactions, such as for medical purpose. But there are 
also ``global views" that should 
be noticed. For example, The possible configurations of folded proteins are 
enormous, while that can be observed in living form is rather limited. 
These protein structures generally can be described as belonging to a limit number 
of families. In each family, ignoring the details, the proteins possess 
similar overall conformations, and in many cases the structures show regular 
forms or approximate symmetry.\cite{r1,r2,r3,r4,r5,r6} Another example is that single 
domain proteins was observed only within a certain range of sequence length: the 
number of amino acid residues in single domain proteins seldom exceeds 
200. Larger proteins usually fold into multi-domains native states.\cite{r6}

With the accumulation of knowledge about the structures and functions 
of proteins, it was found that many proteins of similar structures pursue 
complete different functions, while proteins with different tertiary structures 
may perform similar functions. These suggested that to understand the protein folding
problem physically, one should first get to know the properties of protein 
structures.\cite{r7} Based on the the concepts from the physics of spin glass, 
study shows that to fold efficiently, proteins require a specially shaped 
energy landscape resembling a funnel. A heteropolymer with a completely 
random sequence generically possess a rugged energy landscape without a 
funnel.\cite{r8,r9} Goldstein {\it et al}\cite{r10,r11} have worked on 
optimizing energy functions for protein structure prediction. They found 
that some structures are more optimizable  than others, i.e., there exist 
structures for which the funneled energy landscape can be obtained within 
a wide range of interaction parameters, while for some other structures the 
 parameters for fast folding are much more restricted. The funneled landscape 
 theory argued that the interactions in the folded structure must act in 
 concert more effectively than expected in the most random cases.\cite{r12} 
 Accordingly, compared with most other structures, the superiority of highly optimizable structure 
 should be that its geometric arrangement permit more sequences to reach the 
 concert interaction states.

Other studies on the thermodynamic of lattice protein models also support the 
above idea.\cite{r13,r14,r15,r16} In the lattice HP models, a protein is 
represented by a self avoiding chain of beads placed on a discrete lattice 
with two types of beads: the Polar (P) and the Hydrophobic (H). A sequence 
is specified by a choice of monomer type at each position on the chain $\{x_i\}$. 
Where $x_i$ could be either H- or P-type, and $i$ is a monomer index. A structure 
is specified by a set of coordinates for all the monomers $\{r_i\}$. The energy 
has the form:

$$H=\sum_{i<j} E_{x_ix_j}{\triangle}(r_i-r_j)$$
where ${\triangle}(r_i-r_j)=1$ when $r_i$ and $r_j$ are adjoining lattice 
sites while they are not adjacent along the sequence, and 
${\triangle}(r_i-r_j)=0$ in other cases. Interaction parameter $E_{x_ix_j}$
differ according to the contact type HH, HP, or PP. Given the interaction 
parameters, it is possible to find out the ground state structure(s) of each 
sequence. Study shows that structures differ markedly in their tendency to 
be chosen by sequences as their unique ground states. The number of sequences 
which choice the structure as unique ground state is called the 
{\bf Designability} of this structure. It was argued that only highly designable 
structures are thermodynamically stable and stable against mutation, and 
thus can be chosen by nature to fulfill the duty of life.\cite{r13} 
Though interaction parameters used may differ strongly in different studies, 
the mostly designed structures do not depend strongly on the detail of 
interactions.\cite{r13,r15,r16} 

From above discussion we see that it should be essential to investigate the 
protein folding problem from structural point of view. To see the 
problem more clearly, we take square lattice HP model as an example. The 
total number of the most compact structures of 36 beads chain
is 57337.\cite{r14} Consider 36 beads {\it homopolymer} 
with interaction parameter $E_{x_ix_j}=E_0<0$. All the 57337 structures 
give the same energy when one such homopolymer fold onto each of them. 
Therefore the folded energy can not be used to distinguish the compact structures from 
each other. The essential here is that of discrimination, or characterization: 
give ways to tell how and why structures differ from each 
other. Nature's way to break the symmetry is to replace homopolymer with 
heteropolymer. From this point of view, the success of lattice protein model 
is that it help to reveal this secret of nature.

Studies focusing on the properties of protein structures is still 
lack,\cite{r17} in spite of some recent elaborations in this 
direction.\cite{ar1,ar2,ar3} In this article we present 
one way to break the symmetry, to distinguish the compact structures of 
lattice model without explicitly considering concrete interaction form. 
However, since only compact structures are considered here, 
an loose constraint is actually set on interactions: interactions under which 
compact structures are preferred as ground energy states. 
The method gives a number called {\bf partition number} ({\bf Partnum}) to 
each compact structure during a simple process. The Partnums 
of structures differ strongly, so giving one way to distinguish them from each 
other. 

In the following section we will give the detail of the method, and 
compare the Partnum with designability. The statistical 
properties of Partnums  are discussed in section {\bf II}. The last section is for some remarks.

\section{The definition and interpretation of Partnum}

    It is easy to find out all the compact structures of certain chain length with 
computer.\cite{ar4} Take 9 beads chain as an example. The search is 
self avoiding and restricted to the $3{\times}3$ square lattice shown in Fig.1(A), and the resulting structures should not be 
related by rotation or reflection symmetry. As a result, there are only three 
starting points, $(0,0)$, $(0,1)$ and $(1,1)$, for the search of structures. 
To find the structures start at $(0,0)$, the first step is to go to $(1,0)$. 
This is the only choice, because $(0,1)$ is a symmetric point of 
$(1,0)$. We give all the structures following this step a 
number $p_1=ln(1)$. Now go to the next site. There are two possible choices: 
$(2,0)$ or $(1,1)$. Since the walk is self avoiding and 
restricted to the $3{\times}3$ lattices, the walk following certain choice
may fail to extend to 9 beads length. The choice that will reach to 9 beads length
 is called acceptable. Suppose that both $(2,0)$ and $(1,1)$
are acceptable. Then each compact structure which will be generated 
following $(0,0) \longrightarrow (1,0) \longrightarrow (2,0)$ or $(0,0) 
\longrightarrow (1,0) \longrightarrow (1,1)$ is given a number 
$ln(1/2)$. Generally speaking, restricting to $3\times 3$ lattice and beginning 
at a starting point, there are totally $8$ steps to finish a self avoiding walk. Each step 
is given one number according to the following rule: if the $i$-th 
step has totally $C$ acceptable choices not being symmetrically 
related, then the step is given a number called {\bf partnum of $i$-th step} 
$p_i=ln(1/C)$. For 2D square lattice, the largest possible choice $C_0$ is 3.
 
Adding all the $8$ numbers and then dividing the sum by $8$, we 
get the Partnum $P1$. Here the structure is actually oriented 
. The consideration of oriented walk is reasonable in the 
case of protein structures, because the native protein structure would become 
unstable if the sequence is reversed, and also protein in life are produced 
successively from one end to another. However, if one consider the start and end
 reversal of the walk as a symmetric operation, then one oriented walk and its reverse 
together correspond to a structure that is not related with the 
direction. In the follows, {\it oriented walk} and 
{\it non-oriented structures} are used to distinguish the two different
ways of viewing structure, and the Partnums corresponding to them are denoted as 
$P1$ and $P2$, respectively. However, when it is no need to distinguish them, 
simply {\it structure} is used and the Partnum is denoted as $P$. 
For the non-oriented structure, the Partnum can be define as: $P2=P1(1)+P1(2)$, 
where $P1(1)$ is the Partnum of one of two oriented walks and $P1(2)$ is 
that of its reverse. 

The Partnums of structures of other chain length can be obtained similarly.

Since the original motivation of developing the Partnums of structures is 
to account for the difference of Designability of structures, in Fig.2 we give the plot 
of Designability against Partnum of orientd structures on $5\times5$ lattice (the 
interaction parameters for calculating Designability is the same as used in Ref.
\cite{r13}). There is not strict correspondence between Designability and Partnum. 
However the linear fit of the data revealed that 
Designability tends to increase with the increase of Partnum (see Fig.2). The same thing 
happens for other sequence length. In the case of $6\times6$ lattice, the 
structure with highest Designability\cite{r13,r15} possess the second largest Partnum ($P2$).
 
According to Fig.1(B), an oriented walk corresponds to one
 path from the root to the top leave of the hierarchical tree. The 
 value of Partnum of the structure is determined by the frequency of the 
 path being disturbed by branches. If the path of a walk meet with 
fewer branches, the Partnum would be larger. This can be 
 compared with the conclusion in Ref. \cite{r16}. In Ref. \cite{r16} a simple version of 
 HP model of protein is employed. A walk is reduced to a string of $0$s 
 and $1$s, which represent the surface and core site respectively, as the 
 backbone is traced. Each walk is therefore associated with a point in 
 a high dimensional space. Sequences are represented by strings of their 
 hydrophobicity and thus can be mapped into the same space. It was found that 
 walks far away from other walks in the high dimensional space 
 are highly designable and thermodynamically stable. For this reason, highly 
designable structures are called {\it atypical} in Ref. \cite{r16}. Here the structures 
 with large Partnum can also be called {\it atypical} (atypical average local freedom) since these structures 
 correspond to paths on the hierarchical tree with fewer branches.

In an analog to the suggestion that nature selected out only highly designable structures, 
we assume that there exists a random process which selects out only the structures 
with the largest Partnum. It is interesting to see what this assumption will result in.

For concise we assume a critical Partnum $P_c$, so that only a small portion of 
oriented walks for which $P1>P_c$ can be selected out.
Two oriented walks are called $n$-level similar if their first $n-1$ steps are 
along the same path, and they branched at the $n$-th steps. 
Suppose $s_1$ is among the structures with the most highest Partnum satisfying 
$P1(s_1)>P_c$. This means that there are few branches along the 
path of $s_1$. As a result, it is difficult to find walks which show high 
level similarity to $s_1$. But if there do exist such walks, 
these walks should have high possibility to be selected out. For example, if
$s_2$ is $N-1$ level similar with $s_1$, $N$ being the chain length, then 
$P1(s_2)=P1(s_1)>P_c$. More generally, let $n_{12}$ being the similarity level 
between $s_2$ and $s_1$. We know that $P1(s_2)=P1(s_1)-(1-{\frac{n_{12}}{N-1}}ln(C_0))$, 
$C_0=3$ being the maximal possible choices per step during the search of structures.
According to this expression, the more similar $s_2$ is to $s_1$, the more possible it is
to be selected out.

Assuming that $s_3$ is another walks with $P1(s_3)>P_c$, but it is dissimilar to
$s_1$. From above discussion we know that there are two families, all the members 
of which are selected out. Within each family, the similarity level of two walks 
is much higher than $n_{13}$, while any two walks from different
families are dissimilar from each other, and the similarity level is $n_{13}$.  
We thus come to the conclusion that the selected walks belong to 
separate families. walks within each family are similar, while walks 
belonging to different families are dissimilar.

For the non-oriented structures, there is no the convenience of the hierarchical 
tree to discuss their properties. But it is believable that the above result 
be kept once similarity between structures is properly defined. 
This is the case for the classification of real protein structures, where 
more or less arbitrary criteria\cite{r1,r2,r3,r18,r19,r20,r21} are used to define the 
similarity between protein structures and to classify structure into families, 
superfamilies , folds, and so on.

\section{The statistical properties of Partnums}
    Natural single domain proteins exist only within a limit range of sequence 
length. By both theoretical and numerical studies it is showed in Ref. \cite{r22} 
that the stability of folded sequences against mutation decrease with the 
increase of chain length. In that follows the dependence of the statistical 
properties of Partnum on chain length will be discussed. We will show how some 
structural properties are determined by general statistical principle. 

The density distribution of $P2$ are shown in Fig.4. Things are similar for $P1$. 
In both cases, visually the distribution becomes more and more normal.
Actually it will be shown that the distribution is Gauss distribution in 
the long chain limit. As the first step, however, let's much generally, 
assume that the Partnums of chain length $N$ can be described by 
a density distribution function, $F(P,{v_1,v_2...})$, $v_i$ being the 
moment of $i$-th orders. It is easy to get the average $v_1=<P>$ and variance 
$v_2=\triangle P$. The results of both oriented walk and non-oriented structures are 
shown in Fig.3. 

Fig.3 shows that $<P>$ (both $<P1>$ and $<P2>$) decrease with the increase of chain length. However, 
from the definition of Partnum we know that $<P1>$ ($<P2>$) can not be smaller than 
$-ln3$ ($-2ln3$). So, for either oriented walks or non-oriented structures, 
there must exist $\delta$, so that 
$$lim_{N \longrightarrow \infty}<P>=\delta. $$ 
A similar argument applies to $\triangle P$, where
$$lim_{N\longrightarrow\infty} \triangle P=\epsilon, \epsilon\geq 0.$$

It is known that the total number of compact 
structures $M$ increase exponentially with the increase of chain length $N$: 
$M(N){\sim}(C_{av})^N$, $C_{av}<C_0=3$ being the average number possible choices  
per step for the walks. This gives one way to estimate the value of $C_{av}$ using the 
knowledge of $M(N)$. Fig.5 show the fit of the data $M(N)$ to 
$f(N)=(lnC_{av})N+b$.
The result is $C_{av}=1.397$. Viewing this value of $C_{av}$ as the value 
in long chain limit, we get that ${\delta}=ln(1/C_{av})=-0.3343$ for 
oriented walks, a reasonable estimation (see Fig.3). It should be 
noticed that $C_{av}$ get this way
 is much larger that given by mean field consideration,\cite{ar4,ar5} where 
$C_{av}=C_0/e=1.1$. According to this $C_{av}$, $\delta=-0.099$ for non-oriented walks. 
From Fig.3 we know that this is a value too large to be the long chain limit 
of $<P1>$. So it seems that the mean field 
treatment does not apply to the two dimensional protein model.

With the help of 
central limit theorem, we can argue that the density distribution is Gauss 
distribution in long chain limit, and ${\epsilon}=0$. See follows.

In the space of compact structures, the Partnum $P$ of 
certain structure is the average of the partnums $p_i$ of all the $T$ steps. For 
oriented walks $T$ equals to the chain length subtracted by 1, and for non-oriented 
structures this value should be doubled further. Now divide the $T$ partnums 
into $(T)/n$ groups (suppose $T/n$ is an integer). In each group the $n$ 
members are chosen randomly within the total $T$ numbers. For each 
group we define a new random variable $q_k=\sum_ip_i/n$, $k$ being the group 
index. Since the members in each group are chosen randomly form the total 
$T$ numbers, the $T/n$ newly defined random variable should have the 
same average and variance when $n \longrightarrow \infty$. 
At the same time, since $P=\frac{\sum_kq_k}{(T-1)/n}$, applying the central 
limit theorem,\cite{r23} we know that $P$ is a Gaussian random variable, 
and $\delta P \longrightarrow 0$ when $T/n \longrightarrow \infty$.

From the above discussion we know that, according to Partnums, 
statistically all the compact structures become indistinguishable in long 
chain limit. Recalling the selection rule assumed above, we know that 
it becomes increasingly difficult to select out $atypical$ structures when 
chain length increases. These results show some connection to the work of 
Ejtihadi {\it et al.}.\cite{ar3} With a purely geometrical approach, 
they were able to reduce largely the candidates of structures that can 
be chosen as the ground states of sequences. They found that for the 
case of HP protein model the number of ground state candidates grows 
only as $N^2$, $N$ being the sequence length. While, as pointed out above, 
the total number of compact structures increase exponentially with the increase of 
$N$. So it becomes increasingly difficult to find the ground state 
candidates. This is in accordance with the statistical property of Partnum. 

For fulfilling biology functions, proteins should possess some properties, for 
example fast folding, thermodynamically stable and stable against 
mutation.\cite{r12,r13,r22,r24,r25} 
It was postulated that with the increase of sequence length, the folded 
structures become more and more difficult to possess these properties.\cite{r22} 
Based on the study of Partnum, we propose that this property of 
proteins is determined by the statistical properties of protein 
structures, the detail of interaction having weak influence.

\section{Conclusion Remarks}
Protein structures seem to be a very special class among all the possible 
folded configurations of polypeptide chain. We now know something about {\it how} 
special it is, but little on {\it why} it be so. Ways of characterizing 
folded structures, from whatever point of view, will help to deepen our 
understanding about protein structures.  In this paper, the study on 
Partnum itself is interesting, and more interesting when compared with the 
dynamic and thermodynamic study of proteins. The concept of Partnum is simple 
and can only be applied to lattice model. But the study on it reveals that 
it is possible to investigate protein structures with no consideration of 
interaction detail.

\section*{ACKNOWLEDGMENTS}
\ \ \ \ The authors would like to give thanks to Proff. Wei-Mou Zheng and Proff. 
Bai-Lin Hao for stimulating discussion. We also thank Mr. Guo-Yi Chen for 
many helps on computation.


\newpage

\begin{figure}[p]
\centerline{\epsfxsize=12cm \epsfbox{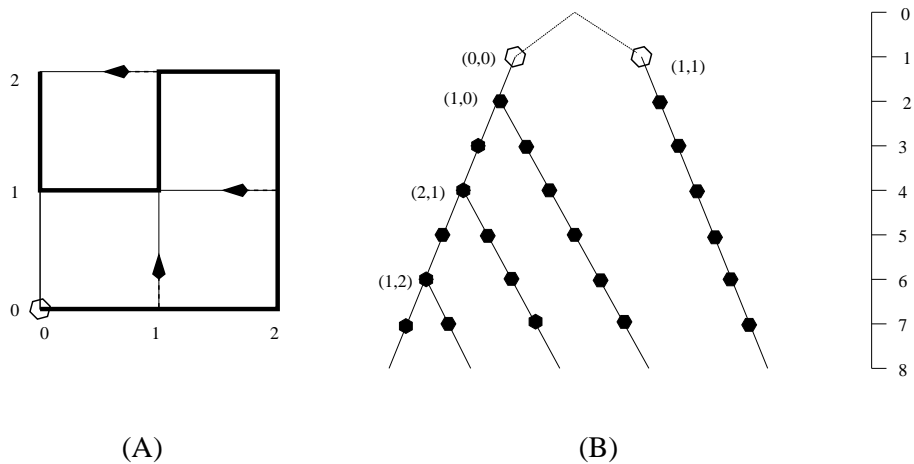}}
\label{fig1}
\caption{ (A):The $3{\times}3$ square lattices used to find out all the compact structures of 
9 beads chain. The bold curve is an oriented walk start at (0,0) and 
end at (0,2). The arrows show that instead of walking along the bold curve, one 
can find other structures in the direction of the arrows. (B): The oriented walks 
and their branching pattern during the search of them. Note that only some 
points show branching on the tree. Others are truncated because they can not extend to 
9 beads length due to the restriction of lattice size and self avoiding. 
The number at the right of the figure show the steps of the search.}
\end{figure}

\newpage

\begin{figure}[p]
\centerline{\epsfxsize=6cm \epsfbox{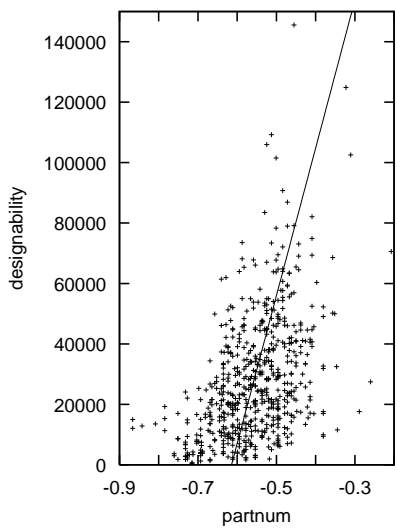}}
\label{fig2}
\caption{Points: Designability against Partnum of non-oriented compact structures 
of chain length 25. Line: the curve of 
$f(x)=ax+b$ with $a=488809$ and $b=300344$. The correlation coefficient 
is $r=0.447$, with totally 621 data points.
}
\end{figure}

\newpage

\begin{figure}[p]
\centerline{\epsfxsize=12cm \epsfbox{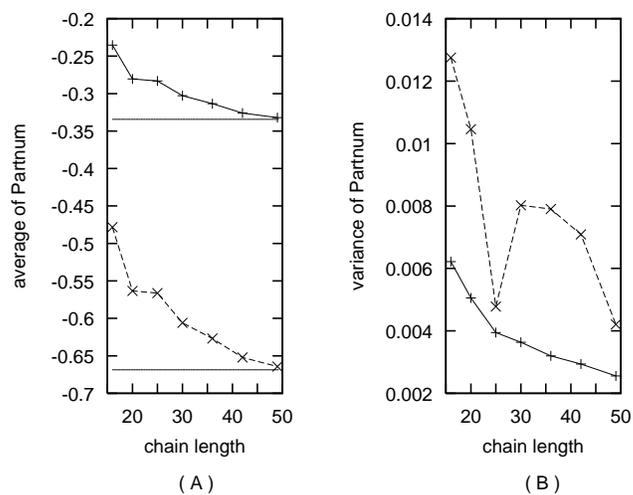}}
\label{fig3}
\caption{ (A): The dependence of the average of Partnums $<P>$ on chain length. 
The upper line-points curve is for the oriented walks, and the lower line-points curve 
is for the non-oriented structures. The upper and lower doted straight lines 
$<P>=-0.3343$ and $<P>=-0.6686$ are the estimated long chain limit of $<P1>$
 and $<P2>$, respectively (see text). (B): The dependence of
the variance of Partnums on chain length.}
\end{figure}

\newpage

\begin{figure}[p]
\centerline{\epsfxsize=12cm \epsfbox{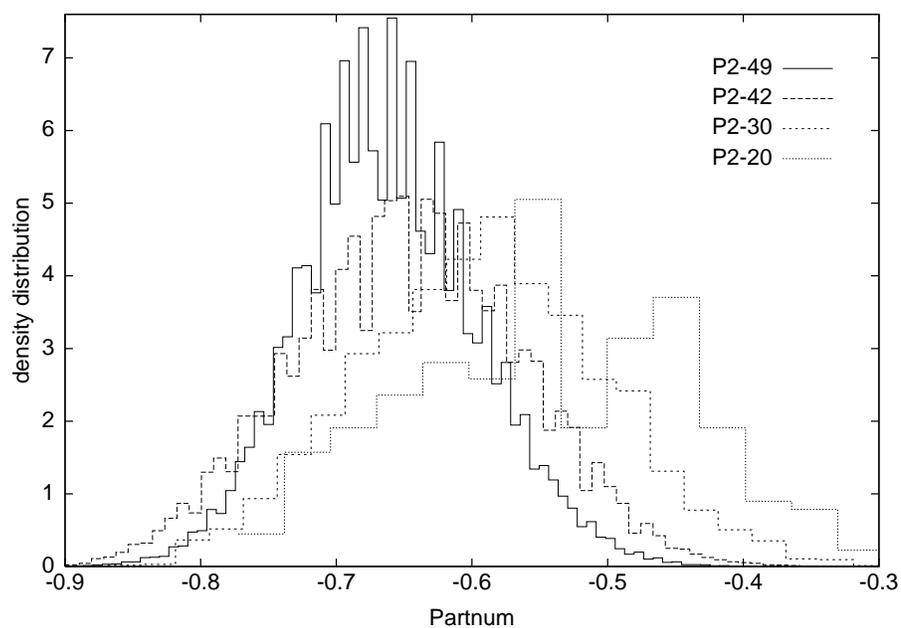}}
\label{fig4}
\caption{The density distributions of Partnums of non-oriented structures under various 
chain length. The number ``49" in ``P2-49", for example, is the chain length. 
The distribution curves are shown in step curve style.}
\end{figure}

\newpage

\begin{figure}[p]
\centerline{\epsfxsize=12cm \epsfbox{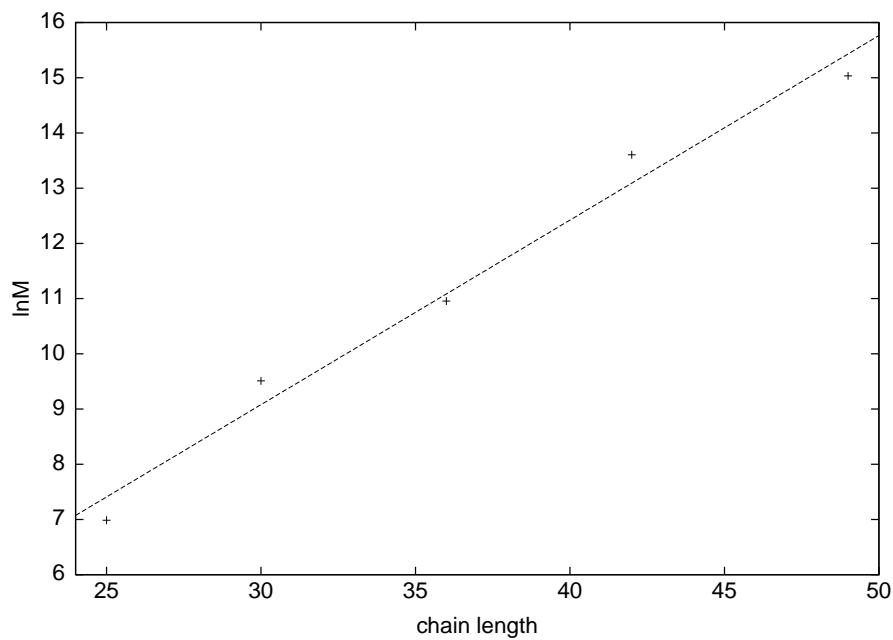}}
\label{fig5}
\caption{Logarithm of the total number of oriented walks versus the chain length. The
line is the fit using $lnM=ln(C_{av}){\times}N+b$, with $C_{av}=1.3969$ 
and $b=-0.9489$. The correlation coefficient is $r=0.99$.
}
\end{figure}

\end{document}